\documentstyle[11pt,aaspp4]{article}
\font\sc=cmr7
\font\sb=cmr10
\def\CIV{C$\,${\sb IV}$\,\lambda$1550}
\def\CII{C$\,${\sb II}$\,\lambda$1334}
\def\SiIIa{Si$\,${\sb II}$\,\lambda$1260}
\def\SiIIb{Si$\,${\sb II}$\,\lambda$1527}
\def\OISiII{O$\,${\sb I}+Si$\,${\sb II}$\,\lambda$1303}
\def\SiIV{Si$\,${\sb IV}$\,\lambda$1403}
\def\SiIVb{Si$\,${\sb IV}$\,\lambda$1394}
\def\AlII{Al$\,${\sb II}$\,\lambda$1671}
\def\AlIII{Al$\,${\sb III}$\,\lambda\lambda$1855,1863}
\def\FeII{Fe$\,${\sb II}$\,\lambda$1608}
\def\MSUN{${\rm M}_\odot$}
\begin{document}
\title{A Proto-Galaxy Candidate at $z=2.7$ 
Discovered by Its Young Stellar Population}
\author{
H.~K.~C.~Yee\altaffilmark{1,2,3},
E.~Ellingson\altaffilmark{3,4},
Jill Bechtold\altaffilmark{5},
R.~G.~Carlberg\altaffilmark{6},
J.-C.~Cuillandre\altaffilmark{3,7}
}

\altaffiltext{1} {Department of Astronomy, University of Toronto, Toronto, 
Ontario M5S 1A7, Canada, Email: hyee@astro.utoronto.ca}
\altaffiltext{2}{Canada-France-Hawaii Telescope, 
P.O.Box 1597, Kamuela, Hawaii 96743}
\altaffiltext{3}{Guest observer, Canada-France-Hawaii Telescope, 
operated jointly by the NRC of Canada, CNRS of France, and the University 
of Hawaii}
\altaffiltext{4}{CASA, University of Colorado, Campus Box 389,
Boulder, CO 80309-0389, Email: e.elling@casa.colorado.edu}
\altaffiltext{5}{Steward Observatory, University of Arizona, Tucson AZ 85721,
Email: jbechtold@as.arizona.edu}
\altaffiltext{6} {Department of Astronomy, University of Toronto, Toronto, 
Ontario M5S 1A7, Canada, Email: carlberg@astro.utoronto.ca}
\altaffiltext{7}{Laboratoire d'Astrophysique de Toulouse
Observatoire Midi-Pyr\'en\'ees, UPS, 14 Av. E. Belin,
31400 Toulouse, France, Email: cuilland@srvdec.obs-mip.fr}

\received{}
\accepted{}

\begin{abstract}
A protogalaxy candidate at $z=2.72$ has been discovered
serendipitously by the CNOC cluster redshift survey.
The candidate is an extremely luminous ($V=20.5$ mag, absolute mag --26)
and well resolved (2$''$$\times$3$''$) disk-like galaxy.
The redshift is identified from  a dozen strong UV absorption lines,
including lines with P-Cygni profiles, which are
indicative of the presence of young O and B stars. 
No emission lines 
are found between 1000 and 2000\AA~(rest), including Ly$\alpha$.
The surface brightness profile of the galaxy fits an exponential
law with a scale length of $\sim$3.5 kpc.
The multi-color photometric data fit the spectral energy distributions
of a stellar population from 400 million years to an arbitrary
young age, dependent on the amount of dust extinction.
However, the presence of a strong P-Cygni profile in \CIV~indicates
that a very substantial component of the stellar population 
must be younger than $\sim$ 10 Myr.
These models predict that this galaxy will evolve into
a bright galaxy of several $L^*$ in brightness.
We can interpret this object as
an early-type galaxy observed within 
about 100 million years of the initial 
burst of star formation which created most of its stellar mass,
producing the extremely high luminosity.
Because of the resolved, regular, and smooth nature of the object, 
it is unlikely that the high luminosity is due to gravitational
lensing.
We estimate the sky density of this type of objects observable
at any one time to be $10^{0\pm1}$ per square degree.

\end{abstract}

\section{Introduction}

The discovery of a primeval, or proto-, galaxy 
 -- loosely defined as a galaxy
in its initial star formation stage --  has long
been one of the most sought-after observational
goals of extragalactic astronomy.
The importance of proto-galaxies (PGs) to our understanding
of the formation of structures and evolution of galaxies is
enormous.
By observing galaxies in their early stages of
evolution, we may be able to delineate the dynamical
processes that lead to and accompany the formation
of galaxies. 
Furthermore, because evolution is slow for galaxies containing
mostly stars older than about 1 Gyr (Tinsley 1972),
it is only by observing the properties of galaxies
during the first Gyr or so of their lifetime that
we will be able to place  strong constraints on
the star formation and chemical evolution
history of galaxies.

It is precisely because no definitive observational data
for PG exist, that over the years PGs have been predicted
to have properties covering the whole imaginable range of 
morphologies, colors, redshift of formation,
and emission-line characteristics
(e.g., see the excellent reviews by Koo 1986 and
Pritchet 1994, and references therein).
In response to the predictions arising from these diverse models
of PGs, many different methods have been suggested and
carried out in their search -- narrow-band imaging from the near-UV
to the infrared searching for Ly$\alpha$ emission 
(e.g., Pritchet \& Hartwick 1990);
 broad-band imaging searches for the Lyman break
(e.g., Steidel \& Hamilton 1993, Djorgovski 1992);
and searching around other known
high-redshift objects discovered by their non-stellar emission,
such as quasars, radio-galaxies, and damped Ly$\alpha$ absorbers
(e.g., Djorgovski 1985, McCarthy 1993, and Wolfe 1995).
Most of these searches have one thing in common: the dependence
of redshift identification on the existence of the 
Ly$\alpha$ emission line.

Many different types of objects have been put forward as
possible PG candidates,
ranging from high-redshift radio galaxies to
Ly$\alpha$ emitters found around quasars and damped Ly$\alpha$
systems.
Most recently, two promising candidates were announced in the literature.
One was the ultra-luminous IRAS galaxy 10214+4724 at $z=2.29$
(Rowan-Robinson et al.~1991).
However, recent observations (Matthew et al.~1994,
Eisenhardt et al.~1996) have clearly shown that gravitational
lensing is the culprit behind the very large apparent luminosity;
the object is consistent with being similar to low-redshift luminous
IRAS galaxies.
Another recent object of interest is Hawaii-167, an unresolved
object at $z=2.33$ with strong UV absorption lines, discovered by
Cowie et al.~(1994) from their $K$-band redshift survey.
Egami et al.~(1996) interpret this object as a buried quasar
with a host galaxy undergoing a strong burst of star formation.
Cowie et al.~(1995) recently reported the discovery of many
galaxies at $1<z<1.6$ with star formation rates in excess of
10 ${\rm M}_\odot$ yr$^{-1}$.
However, these massive star forming galaxies are most likely to
be galaxies with an on-going (if episodic)
 star formation history, and not in their first 
Gyr of their life time.
Hence it appears that a PG candidate that everyone can agree 
on continues to elude searchers.

 From a very simplistic view, an object that is at high redshift,
looks like a galaxy, and is identified by spectral features 
consistent with arising entirely from a large number of young stars
would certainly qualify as a PG candidate.
In this paper, we describe the serendipitous discovery of an
ultra-luminous galaxy at $z=2.72$, identified by its 
absorption-line spectrum
which appears to have all of these attributes.
The object was observed as part of the CNOC (The Canadian Network
for Observational Cosmology) Cluster Redshift
Survey (Carlberg et al. 1994; Yee, Ellingson, \& Carlberg 1996,
hereafter YEC).
The most conservative interpretation of the spectrum is that
it is primarily due to 
young  O, B, and possibly A stars.
With its resolved morphology, 
 galaxy-like surface profile, high redshift,
and extremely high luminosity (about --26 mag at rest 1500\AA), 
this object is consistent with being a $L^*$ galaxy observed within
the first 100 Myr since its initial star burst.
This object appears to fit every definition of a PG.

  In Section 2 we describe the imaging and spectroscopic
observations that led to this discovery.  
Section 3 presents the analysis of the spectra and images.
We discuss the implications of the results in Section 4.
Our conclusions are summarized in Section 5.
In this paper, we use $H_0=75$ km s$^{-1}$ Mpc$^{-1}$, 
and $q_0=0.1$ throughout.

\section{Observations and Reductions}

The CNOC Cluster Redshift Survey (see YEC) was carried out to map the 
velocity field of intermediate redshift EMSS clusters (Gioia et al.~1990)
with the primary goal of determining the global mass-to-light
ratio of galaxy clusters out to large radii.
The project obtained redshifts for $\sim$2600 galaxies with
magnitudes mostly between Gunn $r$=18 and 22 mag.
It was in the process of reducing and analysing this large
data set that the PG candidate was discovered in the field
of the galaxy cluster MS1512+36.
In this section, we briefly describe the relevant information
of the CNOC observations and data leading to the identification
of the PG, and additional follow-up observations.

\subsection{CNOC Imaging and Spectroscopy}

The EMSS cluster MS1512+36 at z=0.373 (see Gioia \&
Luppino 1994) was observed at the 3.6m
Canada-France-Hawaii Telescope (CFHT) on 1993 June,
using the MOS arm of MOS/SIS 
(Multi-Object-Spectrograph/Subarcsecond-Imaging-Spectrograph,
see Le F\'evre et al.~1994).
A detailed description of the observing procedure and
imaging and spectroscopic data reduction techniques for
the survey is presented in YEC.
A complete data catalog for this particular cluster is presented
in Abraham et al.~(1996).

Images of MS1512+36 in Gunn $r$ and $g$ were taken using MOS
in the imaging mode with the 2048$\times$2048 LORAL3 CCD on
1993 June 18.
The scale of the images is 0.3128$''$ per pixel.
The integration time of 900 seconds and seeing of 0.9$''$
provided a 5-$\sigma$ detection limit of about 24.0 mag for
both filters.
Photometry for objects in the field was performed using
the program PPP (Yee 1991, also see YEC).
Calibration to the Gunn system was done using only 3 standard
stars, giving a systematic uncertainty of about 0.07 mag.

Multi-object spectroscopy of galaxies in the field was
obtained using slit masks  designed from the $r$ image.
Two exposures of 60 minutes each were taken on 1993 June 19.
For the CNOC program, we were primarily interested in
the cluster galaxies; hence, band-limiting filters tuned
to the major absorption features of the cluster galaxies
were used to shorten the spectra to increase the
multiplexing capability of MOS.
For MS1512+36 the Z4 filter (see YEC),
optimized for clusters with redshift between 0.37 and 0.45,
was used.
This filter has a spectral range of 4700 to 6300\AA.
The average dispersion of the O300 grism is about 3.45\AA~per pixel,
providing a resolution of $\sim$16\AA~with  a 1.5$''$ slit width. 
The slits were oriented EW.
Spectral reduction 
and cross-correlation techniques used to extract spectra and
obtain galaxy redshifts are described in detail in YEC.

\subsection{The Candidate Object}

Along with cross-correlation, each spectrum was also examined by eye
to double-check the validity of the redshifts and
to inspect visually objects for which no redshift is obtained.
One object, designated spectroscopically as 1512-cB58 
(the 58th spectrum of the B mask in the central field), 
or PPP\#101120,
in the field of the cluster MS1512+36, yielded a relatively high signal-to-noise ratio
spectrum with a large number of strong absorption lines, none of
which correspond to the usual lines seen in optical
spectra of moderate-redshift galaxies.
The discovery spectrum of this object is presented in
Figure 1.

Inspection of the image of the object indicates that
it is clearly a galaxy, with a relatively bright $r$ magnitude of
20.5 and situated only 6$''$ from the brightest cluster galaxy (BCG)
of MS1512+36.
Figure 2 shows the central section of the $V$ image taken
subsequently with SIS (Subarcsecond-Imaging-Spectrograph) at
CFHT (see Section 2.2) with the galaxy cB58 marked.
Enlargements of the PG candidate are shown both as
a gray scale and a contour plot in Figures 3$a$ and 3$b$.

Analysis of the spectrum identified 9 lines in the ultraviolet
corresponding to an absorption system at $z=2.72$.
These lines are marked on the spectrum in Figure 1.
The high redshift, large luminosity, and the resolved nature of
the image all indicate that this is a highly unusual object.

\subsection {SIS Imaging and Spectroscopy}

Because of the remarkable nature of cB58, director's
discretionary time at CFHT was requested in 1995 July to
obtain further images and spectroscopy of the object.
The primary objective was to obtain longer wavelength
coverage, including the vital Ly$\alpha$ region, and
higher quality images.
These new observations were carried out using the
SIS arm of MOS/SIS, which provides tip-tilt correction 
capability and small pixel sampling (0.0869$''$ per pixel) 
to achieve the best possible image quality.

Images in Johnson $I$ and $V$ were obtained on 2 separate
runs (1995 July 5 and 19).
Because the primary purpose for taking the images was for making
slit masks for spectroscopic observations, they are not very deep.
During each night, exposures of 300 and 600 seconds in $I$ and $V$,
respectively, were obtained.
The seeing was 0.64$''$ and 0.73$''$ FWHM, for the averaged $I$
and $V$ image, respectively.
A single standard star field was obtained for each of the nights
(NGC7006 for 1995 July 5 and
NGC 7790 for 1995 July 19; see Christian et al.~1985).
However, because the images from 1995 July 5 were obtained under 
non-photometric conditions, the photometry is calibrated to
that of 1995 July 19.

Spectroscopic observations were made 1995 July 6 using the V150
 grism which gives a dispersion of approximately 4.0\AA~per
 double-binned pixel. 
Two exposures of 40 minutes each were obtained.
A slit of width 0.8$''$ was used, providing a resolution of
about 15\AA.
Due to position angle constraints of other programs being carried
out on the same night, the slit was oriented north-south, along
the minor axis of the galaxy.

The spectrum, shown in Figure 4, was extracted and reduced using 
IRAF\footnote{IRAF is distributed by
the National Optical Astronomy Observatories, which is operated
by AURA Inc.~under contract with NSF.}.
Extraction was performed using variance-weighting, and
the spectrum was wavelength calibrated using vacuum wavelengths of
a HeNeAr arc spectrum. Residuals from the calibration
solution are $\sim$ 1\AA.  The spectrum was flux calibrated
and extinction corrected using an observation of the spectroscopic 
standard BD +40 4032. The flux calibration is estimated to
be accurate to about 15\% at the ends of the spectrum, and
about 10\% in the center. However, since the spectrum
was taken with a very narrow slit, these uncertainties
are lower limits with regard to true spectrophotometry.

\section{Analysis}

\subsection{Spectroscopy}
The absorption line features in both spectra were
identified and measured using the method described in Bechtold (1994).
The observed wavelengths of the
lines, along with their rest equivalent width, identification,
and redshift,  are presented
in Tables 1 and 2 for the MOS and SIS spectrum, respectively.
The lines are marked on Figures 1 and 4.

To determine the redshift, the \CIV~line is not used because
of its P-Cygni profile (see below) .
Using the 8 lines from the MOS spectrum, an average redshift
of 2.7233 with a dispersion of 0.0014 is obtained.
The rms of the mean is 0.0005.
With 9 lines from the SIS spectrum, (discarding also
the \AlIII~lines which are in the atmospheric absorption A band), a 
redshift of 2.7229 with a dispersion of 0.0027 is found.
The redshifts determined from the two spectra are entirely
consistent within their errors.
We also determined the redshift using an artificial template
with the absorption lines
(excluding \CIV)
 represented as delta functions.
A redshift of 2.7233 is derived from the SIS spectrum, 
in complete agreement with that obtained by
measuring individual lines.

  In both spectra, the centroid of the \CIV~lines have a
redshift about 3.5$\sigma$ lower than that of the mean
redshift of the other lines.
This is not the effect of blended lines, as using
either 1548 or 1551\AA~as the rest wavelength does not
alleviate the discrepancy.
This shift is most likely due to 
the line having a P-Cygni profile, which
is apparent in the higher signal-to-noise ratio
 MOS spectrum (Figure 1).
The MOS spectrum also shows significant variations in
the spectral line widths.
For lines that are not blended, most notable are
the differences between \CII~and \SiIIb,
which are interstellar lines, and \SiIV~and \AlII,
which could have a dominant stellar component.

\subsection{Imaging}

The magnitudes of cB58 in the 4 filters and their estimated
uncertainties are listed in Table 3.
Also listed are the magnitudes corrected to the AB system (using
the calibrations of Fukugita, Shimasaku, \& Ichikawa 1995),
so that each band has an identical zero point.
Because of the crowded nature of the field immediately
surrounding the PG, care was taken to derive the
correct relative colors of the object.
For each filter, an aperture of 2.2$''$ diameter centered on
the object is used as the object aperture.
This size, while large compared to the seeing disk, is
small enough to avoid contamination from the cD galaxy and
also to ensure the best signal-to-noise ratio possible.
The outside diameter of the sky aperture is 
chosen to be 9$''$ in order to
avoid the center of the cD galaxy.
The 2.2$''$ aperture magnitudes are then corrected to 
total magnitudes using the isophotal surface
photometry out to 
24.1 mag arcsec$^2$ -- essentially
the faintest contour of Figure 3$b$.
The correction amounts to --0.24 mag.
It is not possible to perform photometry to fainter
isophots without significant uncertainty being introduced by the
halo of the cD galaxy.
The uncertainties listed in Table 3 include contributions from both
calibration and sky level estimates.
The former ranges from 0.07 to 0.1 mag, 
and the latter contribute a similar amount mainly due to the relatively
large uncertainty produced by the proximity of the cD galaxy.
At $V$, which is approximately rest 1500\AA, the absolute
magnitude is --26.0, making this the most luminous non-AGN galaxy
ever observed.

The image of the galaxy shows a relatively flat central region
with no indication of a central unresolved source (see Figure 3).
We can set a limit on the existence of a possible point source by
subtracting the brightest point-spread function (PSF) allowed
which does not create 
a central dip in the light distribution of the galaxy.
For the summed $V$ image we obtain an upper 
limit of 22 mag, or
about $<$25\% of the total galaxy light.
However, the lack of observable Ly$\alpha$ or other
emission or a color gradient (see below) make it unlikely that even this 
amount of light can arise from an AGN or quasar component.

Profiles of the galaxy in $V$ and $I$ are obtained using
the SIS images, which have better seeing and better sampled
pixels than the MOS images.
To increase the signal-to-noise ratio, the data from the
2 nights are summed, and the pixels are binned 2$\times$2
(which still provides a FWHM of 3.5 or more pixels).
The profiles are computed using an isophotal method.
Because of the proximity of the BCG of the foreground
cluster, the profile is measured only to relatively bright
magnitudes.

Isophotal contours
separated by a factor of $\sqrt 2$ in
surface brightness are fitted to ellipses.
The fitting shows that cB58 has a remarkably regular morphology,
confirming the impression from the contour plot.
There is essentially no change in the position angles of the
ellipses (varying by less than the measurement precision of
a few degrees), nor is there evidence 
of shifting in the centroids of the ellipses
with increasing radius
(constant within 0.15$''$).
The only significant change with radius is the ellipticity
($1-b/a$, where $a$ and $b$ are the semi-major and -minor
axis, respectively)
which varies from 0.5 at a radius of 1.2$''$
for the major axis
(at $\sim$ 4 $\times$ seeing disk) to 0.34 at the edge of the
galaxy (major axis radius 1.8$''$).
This change as a function of radius is not due to seeing,
as the galaxy isophots become rounder at larger radii.
This may be an indication of a more spheroidal system surrounding
an inclined disk.

The $V$ and $I$ surface brightness
profiles are presented in Figure 5.
The radius plotted is that
of a circle having the same area as the ellipse (i.e., the
harmonic mean of the major and minor axes).
Also plotted is the $V$ profile of the 
PSF from a star 35.1$''$ from the galaxy, 
shifted to show the maximum possible PSF that can be subtracted
from the PG without leaving a negative gradient towards the center.
The $I$ band profile does not extend as far as the $V$ band
because of larger contamination from the cD galaxy in the
redder band.

The profiles are fitted to both an exponential disk and
a de Vaucouleur law, using data outside a
0.5$''$ radius, avoiding the central part which is smeared
significantly by the seeing disk.
The exponential fits for both the $V$ and $I$ profiles are
shown as dashed lines in Figure 5,  while the de Vaucouleur
fit for the $V$ profile, plotted versus $r^{1/4}$,
 is shown in Figure 6.
Using the $V$ band image, for which we have more extensive data, the
exponential fit is clearly superior, with a reduced $\chi^2$ of 0.9,
five times smaller than the reduced $\chi^2$ of 5.0 from 
the de Vaucouleur fit.
However, we note that if we also eliminate the third point from
the center in the
$V$ profile (which may still be contaminated by PSF smearing),
the de Vaucouleur law is quite acceptable.
Hence, although the exponential disk is a better fit, without
higher resolution HST images, we cannot rule out that the 
galaxy profile is also consistent with a de Vaucouleur law.

The best fitting exponential law has a scale length of
$r_o$=0.37$''$, and a central surface brightness $u_V(0)$=20.4 mag
arcsec$^{-2}$.
The fitted $r_o$ translates to 2.8 kpc at $z=2.72$.
Assuming it is a circular disk, the deprojected scale length
is $\sim$3.5 kpc, typical of present-day luminous disk
galaxies.
The $I$ profile gives an identical scaling length for
the exponential fit with $r_o=0.36''$,
indicating that there is no observable color gradient.
Dividing the 2-dimensional images of $V$ and $I$ band pixel by
pixel confirms the lack of significant color gradients or
patchy color differences.

\section{Discussion}
\subsection {cB58 as a Protogalaxy}

The extended nature, regular morphology, and the lack of a 
significant point source or emission lines
rule out the possibility that cB58 is a quasar or a  quasar-like
object.
Furthermore, the lack of color gradients and
the fact that the absorption lines
appear across the whole spatial extent of the spectral images
allow us to eliminate the possibility 
that we are observing the continuum of a background quasar
with the absorption lines being produced by an extended
(foreground) galaxy at $z=2.7$.
Hence, both the continuum and the absorption lines must
arise from a single object.

The strong continuum and absorption lines can be most conservatively
be interpreted as arising from early type O, B, and  possibly A stars.
The spectrum is best compared to the IUE star library in 
Fanelli et al.~(1992) and Kinney et al.~(1993), and the 
IUE spectral atlas of galaxies in Kinney et al.
In particular, the spectrum bears a striking resemblance 
to IUE spectra of
starbursting galaxies such as NGC7552 and others in
Kinney et al.
However, the ultra-violet continuum of cB58 is many hundred 
times brighter than those of nearby starburst galaxies.

The absorption features in the spectrum can be attributed to
both stars and the interstellar medium (e.g., see Kinney et al.~1993;
and Leitherer, Robert, \& Heckman 1995, here after LRH,
 for a detailed discussion).
Strong stellar lines include \SiIV, \CIV,
and N$\,${\sb IV}$\,\lambda$1720
from O stars, and \AlII~and \AlIII~from A stars.
Dominant interstellar lines include \SiIIa, \OISiII,
\CII, \SiIIb, \CIV, \FeII, and \AlII.
Massive hot stars develop strong stellar winds with velocities
up to 2000 to 3000 km s$^{-1}$.
Hence, it is expected that lines predominantly arising from hot
stars are broader than those from the interstellar medium.
Although our spectral resolution is relatively poor ($\sim 900$ km/sec),
the widths of the strong stellar lines are consistent with being produced
by massive hot stars. 
For example, \SiIV~and \CIV, strong lines
associated with O stars,  are significantly
broader than \CII~and \SiIIb, lines that
are attributed primarily to the interstellar medium.
In both spectra both the line shape and the shifted centroid of 
the \CIV~line are strong evidence for the classic P-Cygni profile
expected from early O stars.
The lack of an emission line at Ly$\alpha$ is consistent
with starburst galaxies and HII regions seen locally
(Hartman et al.~1988).

The morphology and the exponential profile of the optical
image indicate that the object is galaxy-like, with
a linear size of $\sim$25 kpc.
The scale length of $\sim$3.5 kpc is entirely consistent with
typical $L^*$ galaxies.
The smooth light distribution, the lack of measurable color gradients
or significant morphological distortion,
 and the extended nature of the 
absorption line in the long-slit spectrum all indicate that
the young stars are distributed relatively uniformly over the
whole object, and that the starburst activity is not localized.

The high luminosity and global nature of the starburst
activity argue that this galaxy most likely has 
very recently undergone a starburst which is massive enough to be
the initial burst associated with the formation of this galaxy.
In the following sections, we examine in more quantitative
detail this interpretation and its implications.

\subsection {Models }

Constructing detailed and definitive
models of the star formation history of a galaxy is difficult,
even at such a young age.
There are many uncertainties in input parameters
for this type of modeling; e.g., initial mass function (IMF),
metalicities, and the nature of dust.
These uncertainties are greatly accentuated because
the PG is being observed at a time when the universe is
expected to be very different from the present.
Nevertheless, the current data, though limited,
allow us to set interesting limits on some of the parameters.

For any star forming galaxy,
three basic components are usually required for modeling:
an existing population of stars, on-going
star formation, and dust.
Each of these components has many parameters which
are not well constrained.
And often, depending on the data on hand, some important
parameters are degenerate. 
For instance, it is often not possible to separate the effects
of stellar age and dust using colors.

For first order modeling of the stellar population of the galaxy, 
we compare the broadband photometry with the spectral energy
distributions (SEDs) from  GISSEL models of 
Bruzual \& Charlot (1993).
The range of models are then investigated in more detail
in conjunction with the spectroscopic data.
It should be noted that such modeling can only be considered
as approximate, as it is certain that
young galaxies at these redshifts will not have the
near-solar metal abundance on which these models are based.

We have not used the continuum of the SIS spectrum for SED model
fitting because of uncertainties in the fluxing.
However, we note that the  SIS spectrum agrees with the
$g$, $V$, and $r$ photometry very well, although it differs
by about 25\% (too bright) at $I$. 
The discrepancy in $I$ is  probably the result of light loss
in the calibration standard star due to the narrow slit used;
such a loss may have been much less severe
for the resolved galaxy.
The accuracy of the photometry is verified by comparing data
from 3 cluster galaxies (at $z=0.373$)
to the SEDs of evolved galaxies 
from Bruzual \& Charlot (1993).

For dust extinction, we use a slightly modified version of 
the average LMC extinction curve from Fitzpatrick (1986).
We have interpolated over the 2175\AA~hump in the LMC
extinction, based on the fact that in 
red end of the SIS spectrum, which are equivalent
to 2200\AA~rest, we see no evidence for the expected
prominent dip due to this feature.
We note that many local star burst galaxies do not
show this bump in the extinction in their IUE
spectra also (Kinney et al.~1993).

Figure 7 shows a comparison of single starburst
models from Charlot \& Bruzual (1993) and the
broad-band colors, which have been corrected for
differing amounts of extinction. 
Unfortunately, extinction and stellar age
are degenerate in modeling the colors.
Assuming that there is no dust, the optical photometry indicates
a single-burst model at an age of about 400 Myr.
This can then be considered as the upper limit of the age
of the star formation, and will be referred to as the
{\it 400-Myr model}.
We note that these results are based on 
the ``standard'' GISSEL default models with 
an upper mass cut-off of 125 \MSUN;
changing this cut-off does not affect
any of the conclusions based on the SED.

As more extinction is added, younger ages fit the data better.
The spectral energy distribution can be fitted by an arbitrarily
young model up to an extinction of about $E(B-V)\sim0.4$, beyond
which the SED is definitely too steep even for the hottest stars.
Figure 7 also shows two models with extinction corrections.
For $E(B-V)=0.18$, the best fitting age is 200 Myr.
As an example of a very young model, using $E(B-V)=0.3$,
the data fit an age of 5 Myr or less.

Within our observational error bars, models for constant star 
formation essentially look
identical to the very young single-burst model at these 
wavelengths.
Figure 8 shows the comparison of constant star forming models
and the data with an extinction correction of $E(B-V)$=0.3.
In the observed optical bands, the SED is not able to
distinguish models of ages between zero and 500 Myr.
Observations in the near-IR bands should be able to set
an upper limit to the age of the constant star formation models.
We note that a SED with less extinction
will not fit any of the constant star formation models.
We shall refer to the dusty young extremes of these models as
the {\it single-burst young model} and the {\it 
constant formation young model},
or generically as the {\it young models}.
We note that for all the models discussed so far, an additional
much older population, if it exists, does not contribute 
significantly to the flux at the observed optical bands.
In the following, we discuss the models in more detail.

The 400-Myr model fits the UV SED very well.
The smooth appearance of the galaxy and the lack of color gradients
are consistent with the low dust content required by this model.
However, this limiting model has a number of serious problems.
First, we note that assuming a standard IMF, 
this model is expected to fade to an evolved galaxy of about
--25.3 mag in $V$ (rest), or about 25$L^*$, a highly
improbably luminosity for a present-day galaxy, even for a BCG.
The ancestor of such a rare
galaxy would have an extremely low probability of being discovered
in our survey (see Section 4.3).
We note that somewhat younger (e.g. 200 Myr) single-burst models
suffer from the same problem. 
The increase in luminosity from the extinction
correction negates most of the compensation from the larger
fading expected from the younger age.
A flatter IMF, or a low-mass cut off which may have been seen
in some low-redshift starburst galaxies (e.g., see Charlot et al.~1993),
could lower the expected evolved rest $V$ band luminosity.
Preliminary $K$ band photometry (to be reported in a future
paper) may be consistent with such a scenario.

A more immediate problem is based on the spectroscopic data.
The 400-Myr model, or any single-burst
models older than about 10 Myr, cannot explain the entire
stellar population.
Our spectra show strong evidence of P-Cygni
profile in \CIV~arising from massive winds from early O stars.
Hence, star formation must be on-going or has stopped
less than 10 Myr ago, and either scenario
 must have a significant effect on the UV continuum.

The young models are favored by the the existence of 
C$\,${\sb IV}~P-Cgyni profile.
The depth of the broad absorption
component of \CIV~and the height of the
emission component are qualitatively
consistent with line profiles of starburst models in LRH.
Since the 3 parameters of age, IMF slope, and upper mass cut-off
are correlated, a number of combinations of these 3 parameters
produce profiles that are similar to the observed one.
In particular,
two continuous formation LRH models strongly resemble our line
profile: one has a Salpeter IMF with an upper mass cut-off of 40 \MSUN, 
and the other has a steeper IMF slope (3.0) with an upper mass cut-off of
80\MSUN. Both have an age of 9 Myr or greater (at which time
the creation and death of massive O stars come into equilibrium).
In Figure 9 we plot the \CIV~profile from the MOS spectrum
overlaid by the model profile having a Salpeter IMF with an upper
mass cut-off of 40\MSUN.
The agreement is excellent.
Models that are significantly younger, with flatter IMFs, or having
more massive upper mass cut-offs, have too large  P-Cygni profiles;
while models that are steeper in the IMF or lower in upper mass
cut-off have too small  profiles.

Similarly, we find that for the single-burst scenario, the
line profile is well-described by models with ages of between 5 to 7
Myr for the whole range of
IMF slopes and upper mass cut-offs in the LRH models.
We note that ``single-burst" star formation is not physical,
but is used as a description for a relatively short
star formation episode that has already stopped at the
time of observation.
As an illustration,
in Figure 9 we also plot a LRH model with the standard IMF and
an upper mass cut-off of 80 \MSUN~at an age of 8 Myr after
a single star burst.
It clearly does not fit the observed profile, indicating
that star formation could not have stopped completely
for  more than 8 Myr.

The constant star formation and single-burst models in theory
can be distinguished by the \SiIVb,$\,$1402 lines.
According LRH, the 5 Myr single-burst models
should have relatively stronger P-Cygni profiles than the
constant star formation models, due to enhancement from
supergiants in the former.
The Si$\,${\sb IV} lines in the PG spectrum are heavily
dominated by interstellar contributions.
There is a possible indication of an emission component
of a P-Cygni profile in the \SiIV~line;  however, there is no
evidence of a blue broad wing from the \SiIVb~line.
This may indicate a preference of the continuous star
formation models over the very young single-burst
models.
Higher quality data are required for a more definitive conclusion.
However, we note that based on a probability argument, it is
not highly plausible that we would observe this galaxies within
such a short time since its cessation of star formation.

Although the C$\,${\sb IV} P-Cygni profile appears
 to be in agreement with some LRH models, a strong caveat is that
they are based on solar metallicity.
Hence, these
results should not be taken too literally, but rather as
indicative.
For lower metalicity massive stars, it is expected that the
 stellar winds will be less strong due to fewer C$^{3+}$ ions.
Hence, qualitatively, the upper cut-off mass could be higher,
the IMF could be flatter, or the age could be younger.
We also note that if there is a significant additional
underlying continuum, the intrinsic P-Cygni profile would
be larger relative to the true young star continuum.

For the young models, we can use the 1500\AA~continuum to
estimate the star formation rate based on the models of LRH.
The observed $V$ (which is equivalent to almost exactly 1500\AA~rest)
 magnitude of 20.6 corresponds to a luminosity of  $4.3 \times 
10^{42}$ erg s$^{-1}$\AA$^{-1}$, assuming no dust correction.
Depending on the upper mass cut-off and the slope
of the IMF, LRH predict 
1500\AA~luminosities between $10^{38.95}$ to 10$^{40.6}$ 
ergs s$^{-1}$ \AA$^{-1}$ 
at an age of 9 Myr
for a constant star formation rate of 1 ${\rm M}_\odot$ yr$^{-1}$.
Taking a value of $10^{40.04}$ ergs s$^{-1}$ \AA$^{-1}$ (for a model
with  standard IMF slope and a 40 \MSUN~cut off),
ignoring extinction for the time being, and
attributing all of the observed $V$ luminosity
to the constant star forming component, we obtain
a star formation rate of $\sim400$ ${\rm M}_\odot$ yr$^{-1}$.
This rate can be considered as a very generous upper
limit for any on-going star formation in a dust-free model.
For the constant star formation young model,
the flux at 1500\AA~has to be corrected for extinction by a factor
of about 12.4, implying a star formation rate of 4700 ${\rm M}_\odot$
yr$^{-1}$.
Since a typical $L^*$ galaxy has a stellar mass of
$\sim 5\times10^{10}{\rm M}_\odot$, 
a galaxy forming stars at this rate will assemble into a relatively
massive 2$L^*$ galaxy of 10$^{11}$ ${\rm M}_\odot$ in stellar
 mass in about 20 Myr.
However, it should be remembered that
the 1500\AA~luminosity is highly dependent on the
parameters used in the star formation model.
For example, with a flatter IMF slope of 1.5 and an upper mass
cut-off of 80 \MSUN, the star formation
rate can be lowered by a factor of 3.6, requiring about 75 Myr
to build up a 10$^{11}$ ${\rm M}_\odot$ galaxy.
This sets the range of the age of the galaxy in a 
constant star formation model to be on the order of 10 to 100 Myr.

Similarly, we can apply the LRH models to
estimate the size of a young single-burst model.
For a single burst of 10$^6$ ${\rm M}_\odot$ yr$^{-1}$,
the  1500\AA~luminosity is
10$^{39.0}$ erg s$^{-1}$\AA$^{-1}$ for the standard LRH model
(Salpeter IMF and upper mass cut-off of 80\MSUN) at an age
of 7 Myr.
Using the observed 1500\AA~luminosity,
the size of the single burst is estimated to be 
4.3$\times 10^{9}$ ${\rm M}_\odot$, assuming no extinction.
The continuum again must be corrected by a factor of 12.4 for
extinction for the young single-burst model,
bringing the burst strength to 5.3$\times 10^{10}$ ${\rm M}_\odot$. 
Thus, for the young single-burst model, the star formation
from the burst alone  amounts to about a $L^*$ galaxy.

In both cases the young models allow the
galaxy to be observable relatively easily
for about 5 to 10 Myr after the end of star formation,
at which time the observed $V$ band photometry will drop 
by about 1 to 2 mag.
Using the extinction required and the K-correction from
GISSEL models and assuming passive evolution from the time of
observation, we estimate that the single-burst formation 
model will produce an evolved galaxy 
about 7 mag fainter in rest $V$, creating a present-day galaxy with
an absolute rest $V$ magnitude of --21.4, or about 1.7$L^*$.
This luminosity is consistent with the mass from the estimates 
of the star formation rate.

Although it appears that the young models afford very reasonable
fit to both photometric and spectroscopic
data, there are, however, a few difficulties that need to
be addressed.
Several problems are related to
the short time scale expected for cB58 to build up a relatively
massive galaxy, especially for the lower end of the age range of
10 to 20 Myr. 
First there is the problem with the creation of metals that we
see so prominently in the spectrum.
Although this can
be explained by postulating a much fainter (at 1500\AA)
 older component that
is completely hidden by the new star formation,
such a model would require the 
present-day descendant of this galaxy to be even more luminous.
Second, the relatively smooth morphology and lack of color gradients
also argue against such a young age.
It is difficult to
imagine such homogeneity in a galaxy after only about 10 Myr
into its initial star formation, since  the typical dynamical time
for a sizable galaxy is about 10$^8$ yrs.
We note that both of these problems can be potentially
alleviated by the flat IMF models, which have star formation
rates such that it will take about a 
100 Myr to assemble a massive galaxy.

The young models are also not consistent
with all of the observational data.
First, although the observed SED 
fits the young models reasonably well,
it is discrepant at the blue end at the 2$\sigma$ level.
A huge on-going star burst produces a very steep blue 
continuum with F$_\lambda ~\sim~  \lambda^{-2.5}$.
The observed continuum is 
essentially flat around 1500\AA~with 
a roll-over around 1300\AA.
The roll-over is seen in both the fluxed SIS spectrum and
the $g-r$ color.
The standard dust extinction curve 
is not able to reproduce the steep
continuum shape required by a young model at the $g$ band
without having to resort to additional absorption beyond 1300\AA.
However, we note that the amount of extra extinction required is within
the uncertainty in the standard extinction curve.
Second, there is some evidence that
that O stars may not completely dominate the
continuum light.
In the spectrum, a strong \AlII~absorption line is seen.
This line likely has a significant stellar component as
it is broader than the pure interstellar lines, 
such as \CII~(see Figure 1).
The \AlII~line could be an indicator of a component
of early A type stars in the continuum.
Additional data in the IR bands may allow us to put
a limit on the size of an older component.

Dust is often thought to be
 associated with forming galaxies or star bursts.
It is likely that our PG candidate does not contain an extraordinary
amount of dust.
The $E(B-V)=0.3$ required by the young models
 is not an outlandish amount
and is consistent with the average
values that are seen towards O star associations in the LMC
(LRH).
The smoothness of the optical images and the lack of significant
color variance argue against significant patchiness due to 
dust.
The lack of Ly$\alpha$ from the HII regions expected from the young
stars requires only a small amount of dust as an explanation.

In summary, our optical photometry is able to limit the
age of the galaxy to younger than $\sim$ 400 Myr, with
the age critically dependent on the extinction correction,
which is limited to $E(B-V)<0.4$.
However, the 400-Myr model is highly unlikely
because of the object's very large luminosity.
Moreover, spectroscopic signatures of a P-Cygni
profile for \CIV~require that there be a significant
number of O stars, and hence a component that is either forming
stars or has an age younger than $\sim$10 Myr
must exist.
Simple standard models over the whole range of ages
all have their respective problems, and
none appears to be
able to explain all the observations so far.
This may be an indication of a non-standard IMF or dust absorption law.
Such a conclusion is not surprising, as there seems little
reason to believe that star formation and dust absorption in a primeval
galaxy should operate in a manner identical to those in nearby galaxies.
The most likely scenario is that this galaxy is somewhere between
50 and 100 Myr in age, contains a small amount of
relatively uniformly distributed dust, and is forming stars
at a high rate, probably with a non-standard IMF.

\subsection {Estimate of Sky Density}

We have shown that we can interpret the galaxy cB58 as
an early-type galaxy observed within 100 Myr of 
its major star formation episode.
Is this object the first of a new, but relatively common,
 class of high-redshift object? 
How likely is such a galaxy be observed serendipitously?
It is of interest to make an order-of-magnitude
estimate of the density of such objects on the sky to
see if it is probable to expect the CNOC cluster redshift
survey to find one.

There are many factors that enter into the detectability
of a PG: the volume density, the luminosity, the
recognizability of the object (i.e., observable spectroscopic
signatures), and the time interval over which these conditions
are favorable.
We can estimate the expected number of such objects using our
knowledge of galaxy density and
the observational parameters of the CNOC survey.
However, it should be noted that many of the parameters used
are highly uncertain.

For an object similar to cB58, there is a relatively small
redshift window in which the object is observable. 
In order for this object to be 
identified unambiguously, the spectral
region that is rich in absorption lines must be in the
observed band, especially when this type of object has
not been observed before.
This spectral region ranges from \SiIIa~to \AlII.
Ly$\alpha$ is not included because of the rapid drop in
the continuum there.
To the red of \AlIII, there are no significant absorption
lines up to well past 2000\AA~rest.
To be able to detect a similar PG, let us assume that
we need to cover the minimum of \CIV~for the
low redshift limit, implying $z\sim2$.
Similarly, we require \CIV~at the red end of the spectrum
for the high redshift limit, implying $z\sim3$.

The CNOC1 redshift survey covers 0.66 square degrees.
The volume between $z=2$ and 3
subtended by 1 sq deg is about  8.4$\times$10$^6$ Mpc$^3$.
Using the present date co-moving density of
galaxies derived from integrating the luminosity
function of Loveday et al.~(1993),
we expect $\sim$1600  galaxies
in the volume sampled by CNOC that are as bright as 2$L^*$,
the expected evolved luminosity of the young models.
If we assume that galaxy formation is uniform between
$z=5$ and 2, then approximately 1/2 of all galaxies
would be formed in the interval of $2<z<3$.
Of these 800 galaxies, about 20\% may be expected to
be early-types with most of their stars formed in
a short-duration large burst, reducing the number to 160.
These forming galaxies will be in an ultra-luminous phase for
only a short period of time.
Assuming this to be $\sim$ 50 Myr,
then about 1/45 of them will be visible at any one time,
bringing the number down to 3.5.
The CNOC survey has an average completeness of about 40\%
between  $r=20$ and 21 mag (mostly due to the 
many low-redshift clusters in the sample).
Hence, the expected number is about 1.5.
The interpretation of this object as a PG {\it can be}
consistent with its serendipitous discovery in the CNOC survey.
Using the above estimates, we might expect to find between 10$^{0\pm1}$
such galaxies per square degree at magnitudes brighter than $r$=21.

Of course, our estimate contains many uncertainties of a
factor much larger than 2 each.
Within the extreme range of models that we discussed, there
may be a factor as large as 100 introduced in the uncertainty
of the sky density.
As an example, if we use the 200-Myr single burst model, the
observable life-time is increased by a factor of about 4,
but it is more than offset by the factor of over 1000  fewer
galaxies expected to be as bright as 10$L^*$.
Other large uncertainties include the volume sampled, which
differs by a factor of 3 between $q_0=0.1$ and 0.5, and
the fraction of $L^*$ galaxies that were formed at $2<z<3$.
For instance, 
if only 10\% of $L^*$ galaxies were formed in this redshift period,
then we would  expect a factor of 5 fewer such objects.
If the bulk of these galaxies were formed at higher redshift,
a fainter spectroscopic survey further out in the red is required.
Also uncertain is the fraction of galaxies that undergo such
a huge burst of star formation.
It should be noted that, conversely, additional discoveries such as this
should allow us to constrain these important parameters.
Our order-of-magnitude estimate here is merely to demonstrate that it is not
entirely unreasonable for a survey such as CNOC to discover such
a PG candidate serendipitously.

\subsection {Possible Effects of Lensing}

The main thrust of the interpretation that cB58 is a galaxy
at the very early stage of its history is its enormous luminosity,
and hence star formation rate.
Gravitational lensing can cause the luminosity of
an object to increase many fold.  
An example of a PG candidate being lensed, and hence
not being as luminous as was first thought, is IRAS10241+4724
(Rowan-Robinson 1991, Broadhurst \& Leh\'ar 1995,
Eisenhardt et al.~1996) at $z$=2.29.
IRAS10214+4724 was first detected in a ground-based image, 
with substantially poorer seeing (1.5$''$) than ours.
However, higher resolution images, both ground-based IR and
HST optical images, show that the object is lensed with an
estimated brightening of about a factor of 30 to 100 
(Eisenhardt et al.~1996).
Another example of a lensed high-redshift galaxy is arc \#384 of
in Abell 2218 with a spectrum similar to cB58 at a
 redshift of 2.51 (Ebbels et al.~1996), discussed in Section 4.5.

Our PG candidate is 6\arcsec~from the cD galaxy in 
MS1512+36, a relatively poor (equivalent to
Abell richness 0) cluster at $z=0.373$
with a velocity dispersion of 690 km s$^{-1}$
(Carlberg et al. 1996). 
The proximity of the BCG
to the protogalaxy candidate requires an
examination of the possible effect of lensing on the
luminosity of the object.
However, we note that even a magnification of a factor of
several in the luminosity does not severely diminish the
interpretation that this object is being observed at a very
luminous phase of its life time.

Large magnification usually occurs in the presence of a 
very strong image shear, as is the case for arcs.  
Our images of the PG under exceptional ground-based seeing
show no sign of distortion, and hence is very unlikely that
it is being strongly sheared, or multiply imaged.
Any lensing similar to that seen in IRAS10241+4724 would have
been easily detected in our $V$ image with 0.65$''$ seeing.
The axes of the galaxy are neither tangential
nor radial with respect to the center of the cD galaxy,
which again argues against the possibility of strong lensing.
In addition, an examination of both our images and the deep image
in Gioia \& Luppino (1994) reveals no sign of other gravitationally
lensed arcs near the cluster.
Based on the axis ratio of the galaxy of $<$2 and the 
very smooth and regular morphology of the object, an upper limit
of a magnification of 2 from weak lensing can be established.
Such a low upper limit does not change the conclusion with regard
to the high luminosity of the galaxy.

The fact that this object is completely resolved in both directions is
the primary argument against gravitational lensing.
However, it is possible, although very
unlikely, that strong magnification can result without shear.
This can occur when there are two subcritical lenses superposed in such a
way that the image lies at a location where the shear of one lens is
at right angles to the shear of the other. 
In particular, if the cD is
off-center and outside the critical radius of the cluster, then in the
directions perpendicular to the cD-cluster center line there will be
two positions of considerable magnification, but with little image shear.
An alternate possibility is that if the cD and cluster potential have
a common center (at least in projection), then a subcritical cluster
without a central cusp has a small zone of large magnification, close
to the cD and with little image shear and no counter images. 
(L.~Williams, private communication).
In both cases a very
precise alignment of two lensing mass profiles possessing a narrow
range of parameters is required. 
At this point, we cannot rule out these possibilities, despite the
low probability.
However, the poorness of the cluster probably precludes a
very large lensing magnification even if the geometry is satisfied.
 
\subsection {Comparison to Other Objects at High $z$}

Most objects found at high redshift have been
discovered as a result of their non-stellar radiation.
These include quasars, high-redshift radio galaxies, and
IRAS galaxies.
They are typically identified by emission lines, especially
Ly$\alpha$,
and stellar spectral signatures have been difficult to observe.
Another major class of objects that are inferred to be at high
redshift are the objects responsible for the quasar absorption lines;
however, little is known about their stellar content.

The object bearing a close spectral resemblance to cB58
is Hawaii-167 at $z=2.33$, discovered by Cowie et al.~(1994).
This object was observed as part of their deep $K$-band survey.
The optical spectrum shows strong absorption lines in the UV,
similar to cB58, and the $I$ band magnitude is nearly identical.
However, the similarities end here, and there are major differences.
Foremost is that Haw167 is unresolved.
Furthermore, it is much redder in the optical: $B-I\sim3$ mag for
Haw167 vs $g-I$=0.7 mag.
(Note that $B$ at $z=2.33$ is identical in rest wavelength to
$g$ at $z=2.72$.)
This color difference is borne out by the drop at about 1800\AA~
rest in the spectrum of Haw167, which is not found in cB58.
Egami et al.~(1995) interpret Haw167 as a dust-enshrouded QSO
with a star burst.
Morphologically cB58 is clearly not a quasar,
and  the difference in color may be due to either
dust or a somewhat older stellar population in Haw167.

Cowie et al.~(1995) have also discovered a significant population of
strong [O$\,${\sb II}]$\lambda$3727 galaxies at $z$ between 1 and 1.7.
They interpret these as massive star forming  galaxies with a
star formation rate of $>10{\rm M}_\odot$ yr$^{-1}$.
It is unlikely that cB58 is a member of this population.
More likely, the objects found by Cowie et al.~represent 
continuous star forming galaxies, (hence, probably later morphological
types), whereas cB58 
may be representative of 
galaxies with a massive initial
burst of star formation creating most of the stellar mass.

Recent observations of arcs in Abell 2218 have discovered
one at $z=2.51$ with an absorption-line spectrum (Ebbels 
et al.~1996).
This arc has a spectrum that  resembles very closely
that of cB58, but with weaker absorption features.
However, the intrinsic luminosity of this lensed galaxy is
several magnitudes fainter than cB58 (Kneib et al.~1996), 
hence, this galaxy is most likely undergoing 
a more minor star-burst.

\section{Summary}

We have serendipitously discovered an excellent protogalaxy 
candidate at $z=2.72$ from the CNOC redshift survey.
Images in four different filters under excellent seeing
show that it is well resolved with a surface profile
consistent with an exponential law with $r_o\sim$ 3.5  kpc.
Modeling the SED sets an upper limit of 400 Myr as the
age of the stellar population, with arbitrary younger
ages possible when the continuum is corrected for dust 
extinction.
The strong absorption-line spectrum, including
the existence of a prominent
P-Cygni profile in \CIV, requires a dominant component with
massive on-going star formation.
Comparison with starburst models indicates that the spectroscopic data
are consistent with either a continuous star formation model
with a star formation rate of several thousand \MSUN~per
year, 
or a 5 to 10 Myr old single-burst 
massive enough to have formed a $L^*$ galaxy.
The lack of morphological peculiarity or a significant color
gradient indicate that the star formation is not localized, but
occurring over the whole galaxy, requiring that the 
the star formation have been 
proceeding for a duration close to 
the dynamical time of the galaxy.
The total luminosity of the this object is extremely large, with
an observed absolute magnitude of $-26$ at 1500\AA.
Assuming either the single-burst young model or the constant star formation
model, we expect this 
object to evolve into a galaxy equivalent to $\sim$2$L^*$.
Because of the the very large star-formation rate,
which can assemble a massive galaxy in 25 to 100 Myr,
we prefer to interpret this object as a 
precursor of an early-type galaxy.
At this point, none of the data indicate  that the high
luminosity arises from gravitational lensing.
Additional observations at other wavelengths, both imaging and
spectroscopic, with higher signal-to-noise ratios and higher
spatial resolution, will allow us to define the star formation
history with more certainty.
 
We estimate the density of such objects on the sky
to be $10^{0\pm1}$ per square degree 
based on this interpretation.
The discovery of one such object in the CNOC cluster redshift survey is
statistically consistent with it being a bright early-type galaxy observed
at a very young stage.
The relatively high sky density expected
for such objects suggests that this
is a member of a relatively common class of objects in the sky.
Additional discoveries and detailed investigations of this 
type of object will provide
immensely important insights and constraints on 
galaxy formation and evolution.

\acknowledgements

We wish to thank Pierre Couturier, the director of CFHT, for
the generous granting of director's discretionary time to
obtain additional imaging and spectroscopy data on this object.
We thank Christian Vanderriest for obtaining some of the SIS
images. 
We would also like to thank Peter Conti, Norm Murray,
Mike Shull, and Liliya Williams for useful discussions.
RGC and HKCY acknowledge NSERC for financial support.
EE acknowledges NASA grant NAG-5-2896 and an NOAO travel
grant.  JB is supported by NSF grant AST-9058510.
HKCY is grateful for the hospitality offered by CFHT,
where part of the manuscript was prepared.


\newpage
\hsize 5.8truein
\raggedbottom
\centerline {
  T{\sb ABLE} 1. Absorption Line List (MOS Spectrum)}
\baselineskip 12pt
\vskip 0.2in
\halign to\hsize{ \hfill #\hfill & \hfill # \hfill & # \hfill &
\hfill # \hfill &   \hfill # \hfill  & # \hfill &
 #\hfill & # \hfill &   \hfill # \hfill   \cr
\noalign{\hrule\vskip 3pt\hrule\vskip 8 pt}\cr
 Wavelength (\AA) & $\pm$ & ~~~ & EW (\AA) & $\pm$ & ~~~ & ID
& ~~~ & $z$ \cr
\noalign{\vskip 5pt\hrule\vskip 7pt}\cr
\noalign{\vskip 5pt}\cr
 4693.9 & 1.4 & ~ & 4.4 & 0.9 & ~ & Si II 1260.4 & ~ &  2.7241   \cr
 4758.0 & 1.4 & ~ & 2.5 & 0.4 & ~ & --- & ~ & --- \cr
 4852.9 & 1.0 & ~ & 5.6 & 0.4 & ~ & OI 1302.2, SiII 1304.4 & ~ & 2.7236 \cr
 4969.0 & 0.9 & ~ & 3.4 & 0.3 & ~ & CII 1334.5  & ~ & 2.7234 \cr
 5189.6 & 0.7 & ~ & 2.8 & 0.2 & ~ & SiIV 1393.8 & ~ & 2.7234 \cr
 5221.2 & 1.3 & ~ & 2.3 & 0.2 & ~ & SiIV 1402.8 & ~ & 2.7204 \cr
 5683.8 & 0.7 & ~ & 2.4 & 0.2 & ~ & SiII 1526.7 & ~ & 2.7229 \cr
 5761.5 & 0.5 & ~ & 5.5 & 0.2 & ~ & CIV 1548.2, 1550.8 & ~ & (2.7183) \cr
 5992.0 & 1.3 & ~ & 2.1 & 0.2 & ~ & FeII 1608.4 & ~ & 2.7253 \cr
 6221.4 & 1.0 & ~ & 2.6 & 0.2 & ~ & AlII 1670.8 & ~ & 2.7236 \cr
\noalign{\vskip 5pt}
\noalign{\vskip 5pt\hrule\vskip 3pt\hrule}\cr
}

\vskip 2truecm
\centerline {
  T{\sb ABLE} 2. Absorption Line List (SIS Spectrum)}
\baselineskip 12pt
\vskip 0.2in
\halign to\hsize{ \hfill #\hfill & \hfill # \hfill & # \hfill &
\hfill # \hfill &   \hfill # \hfill  & # \hfill &
 #\hfill & # \hfill &   \hfill # \hfill   \cr
\noalign{\hrule\vskip 3pt\hrule\vskip 8 pt}\cr
 Wavelength (\AA) & $\pm$ & ~~~ & EW (\AA) & $\pm$ & ~~~ & ID
& ~~~ & $z$ \cr
\noalign{\vskip 5pt\hrule\vskip 7pt}\cr
\noalign{\vskip 5pt}
 4522.7 & 6.1 & ~ & 11.4 & 2.5 & ~ & Ly$\alpha$ 1215.7 & ~ & 2.7203  \cr
 4696.1 & 4.4 & ~ & ~3.0 & 0.9 & ~ & Si II 1260.4 & ~ &      2.7258  \cr
 4848.4 & 3.1 & ~ & ~4.6 & 0.7 & ~ & OI 1302.2, SiII 1304.4 & ~ & 2.7201 \cr
 4963.1 & 1.9 & ~ & ~4.3 & 0.5 & ~ & CII 1334.5  & ~ & 2.7191 \cr
 5112.2 & 2.4 & ~ & ~2.2 & 0.4 & ~ & --- & ~ & --- \cr
 5187.8 & 1.6 & ~ & ~3.2 & 0.4 & ~ & SiIV 1393.8 & ~ & 2.7221 \cr
 5223.3 & 2.2 & ~ & ~2.2 & 0.4 & ~ & SiIV 1402.8 & ~ & 2.7235 \cr
 5688.2 & 1.5 & ~ & ~3.7 & 0.3 & ~ & SiII 1526.7 & ~ & 2.7258 \cr
 5757.5 & 1.1 & ~ & ~6.1 & 0.4 & ~ & CIV 1548.2, 1550.8 & ~ & (2.7157) \cr
 5988.6 & 2.1 & ~ & ~2.6 & 0.3 & ~ & FeII 1608.4 & ~ & 2.7232 \cr
 6225.5 & 1.7 & ~ & ~3.8 & 0.4 & ~ & AlII 1670.8 & ~ & 2.7261 \cr
 6904.8 & 2.3 & ~ & ~5.6 & 0.4 & ~ & AlIII 1854.7, 1862 & ~ &  --- \cr
\noalign{\vskip 5pt}
\noalign{\hrule\vskip 3pt\hrule}\cr
}

\newpage
\hsize 4.truein
\vskip 1truecm
\centerline {
  T{\sc ABLE} 3. Photometry (3$''$ aperture)}
\baselineskip 12pt
\vskip 0.25in
\halign to\hsize{ # \hfill & # \hfill & \hfill # \hfill & # \hfill &
 # \hfill &   \hfill # \hfill   \cr
\noalign{\hrule\vskip 3pt\hrule\vskip 8 pt}\cr
 Band   & ~~~ & Magnitude & $\pm$ & ~~~ & AB Magnitude \cr
\noalign{\vskip 4pt\hrule\vskip 5pt}\cr
\noalign{\vskip 5pt}
 $g$ . . . . .  & & 21.08 & 0.10 & & 21.15 \cr
 $V$ . . . . .  & & 20.64 & 0.12 & & 20.64 \cr
 $r$ . . . . .  & & 20.60 & 0.10 & & 20.41 \cr
 $I$ . . . . .  & & 19.92 & 0.12 & & 20.35 \cr
\noalign{\vskip 5pt}
\noalign{\hrule\vskip 3pt\hrule}\cr
}
\newpage
 \hsize 6.5truein

\figcaption{
Discovery Spectrum of the protogalaxy, plotted in observed
wavelengths. Identified absorption lines are marked.
}   
\figcaption{
Gray scale plot of the central section of SIS image in $V$ of
MS1512+36, showing the center of the cluster and the PG.
The protogalaxy candidate is marked by the letters `PG'.
NE is upper-left.
Intervals on axes are labeled in arcseconds.
}   
\figcaption{
(a) Enlarged gray scale plot of the PG in $V$.
(b) Contour plot of the $V$ band image of the protogalaxy to
the same scale as the gray scale.
Contours are  factors of 1.41 apart beginning at the peak
pixel.  The last contour is 24.1 mag/sq arcsecond.
}
\figcaption{
SIS spectrum of the protogalaxy, plotted in rest wavelengths.
Identified absorption lines are marked. The spectrum has been
smoothed by a box-car filter of 3 pixels in size. 
Note the lack of an emission line at Ly$\alpha$.
}
\figcaption{
$V$ (top panel) and $I$ (bottom panel) surface profiles
 of the protogalaxy.
Exponential law fits to the outer points are shown as dashed-lines.
Also plotted (dot-dashed line) is the PSF from a bright star, scaled
to the maximum brightness that can be subtracted from the PG.
}
\figcaption{
$V$ profile of the PG but plotted versus radius$^{1/4}$.
The best fitting de  Vaucouleur $r^{1/4}$ law 
is shown as a dashed-line.  Note that the exponential law
fit is considerably better.
However, if only the 5 outer points are used, a $r^{1/4}$
law provides an acceptable fit.
}
\figcaption{
Comparison of single-burst star formation models with
$g$, $V$, $r$, and $I$ photometry.
The bottom set of points are the observed fluxes, with the
other two sets of points representing corrected fluxes with
extinction of $E(B-V)$=0.1 and 0.3, middle and top,
respectively.
The horizontal ``error-bars'' represent the half-width
of the filters.
Solid lines from bottom to top  represent 
GISSEL models of 400, 200, and 5 Myr after a single burst .
}
\figcaption{
Comparison of constant star formation models with
$g$, $V$, $r$, and $I$ photometry corrected by an
extinction of $E(B-V)=0.3$.
Solid lines from bottom to top  represent 
GISSEL models of  constant star formation
at ages of 500, 50, and 5 Myr.
}
\figcaption{
Enlargement of the MOS spectrum, shown as the thin sold line,
around the \CIV~region, illustrating the P-Cygni profile.
The other strong absorption line is \SiIIb.
The thick solid line shows the \CIV~profile from the constant star
formation model of Leitherer et al. (1995)~at 9 Myr
with the Salpeter IMF and a 40 \MSUN~upper mass cut-off. 
The agreement is excellent, with the PG spectrum showing
considerably stronger interstellar components.
The thick dashed line shows the model profile at an age of 8 Myr
after a single burst with the Salpeter IMF and an 80 \MSUN~upper
mass cut-off.
The model clearly does not fit as well and 
indicates that the upper limit of the age of a single burst is
about 7 Myr.
}
\begin{figure}[h] \figurenum{1}\plotone{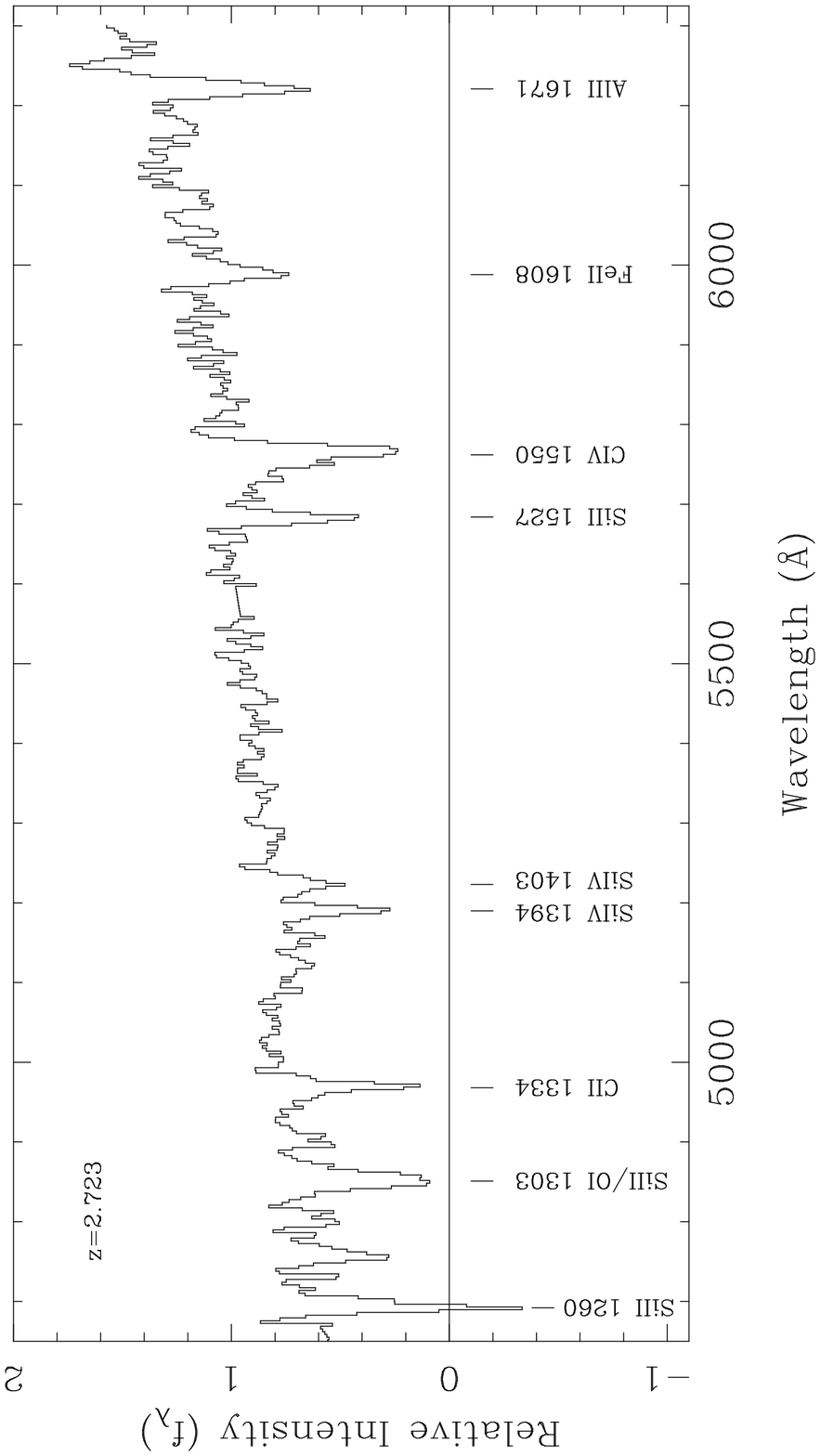} \caption{}\end{figure}
\begin{figure}[h] \figurenum{2}\plotone{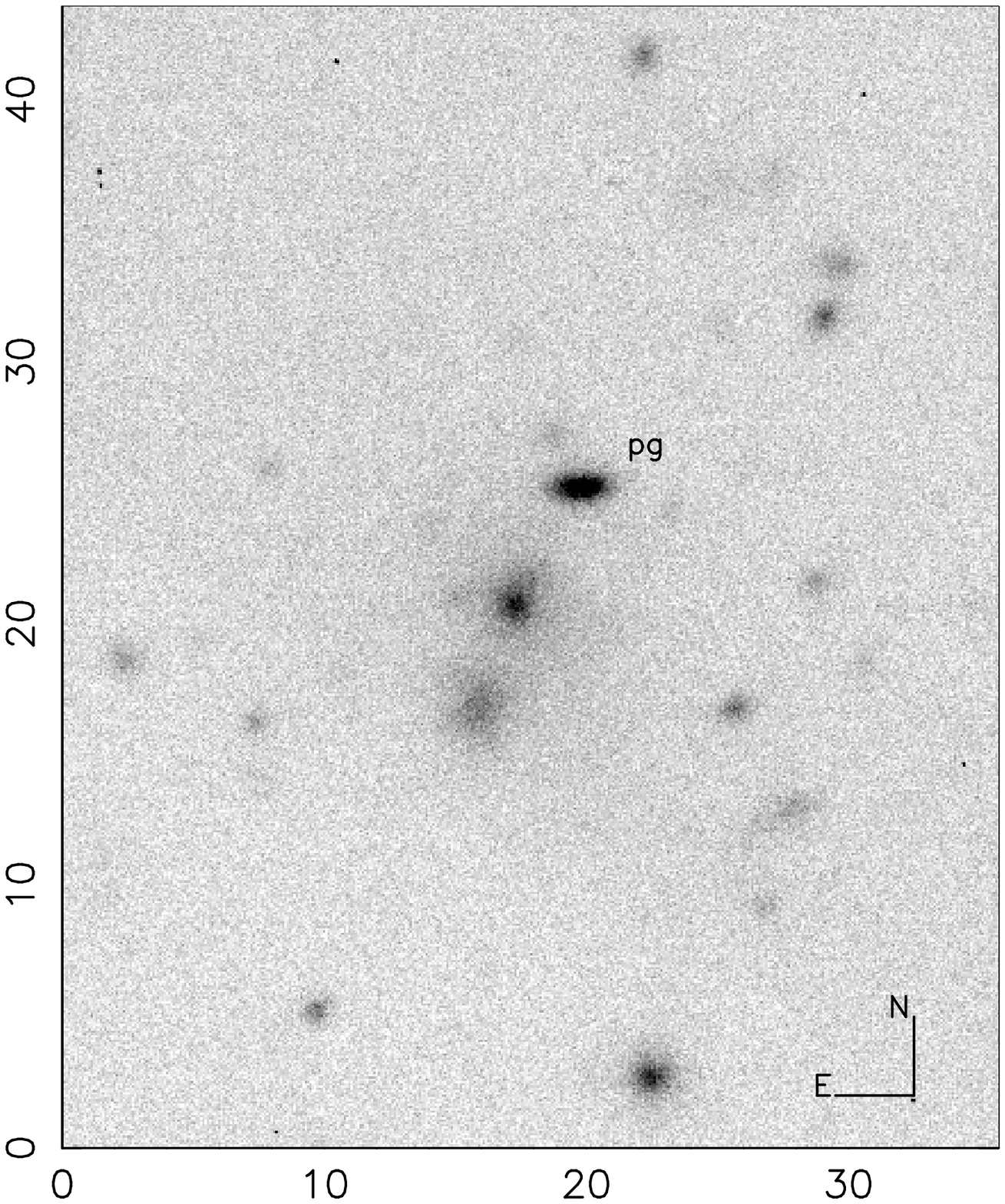} \caption{}\end{figure}
\begin{figure}[h] \figurenum{3}\plotone{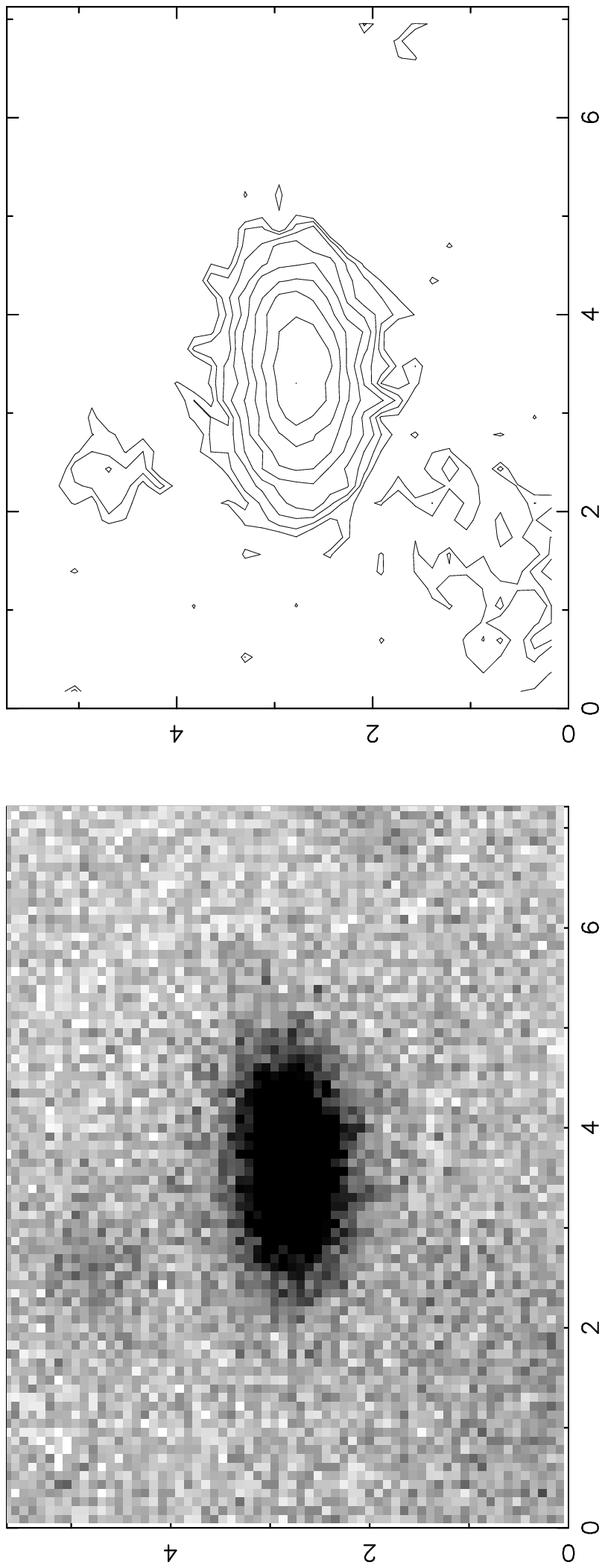} \caption{}\end{figure}
\begin{figure}[h] \figurenum{4}\plotone{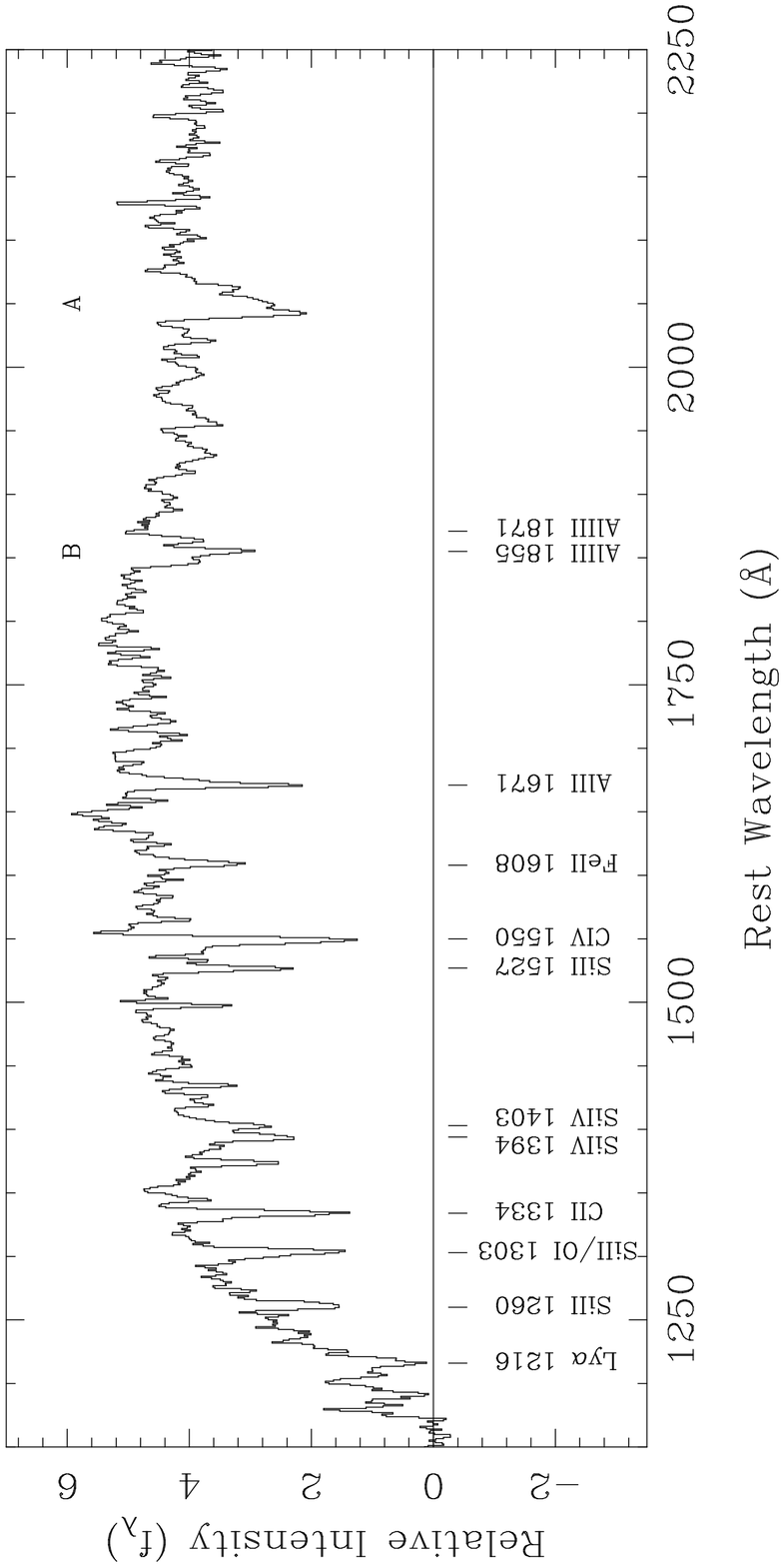} \caption{}\end{figure}
\begin{figure}[h] \figurenum{5}\plotone{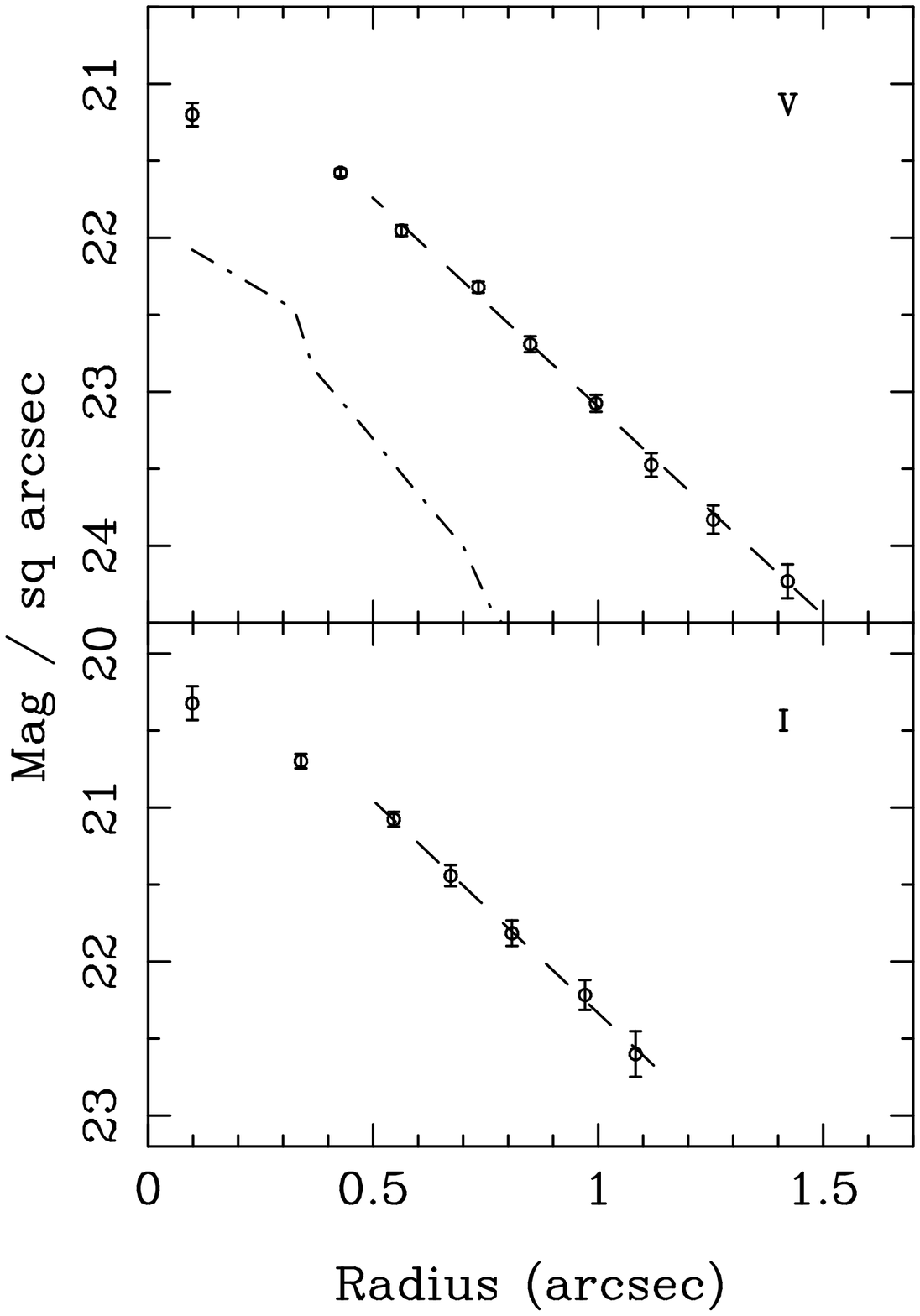} \caption{}\end{figure}
\begin{figure}[h] \figurenum{6}\plotone{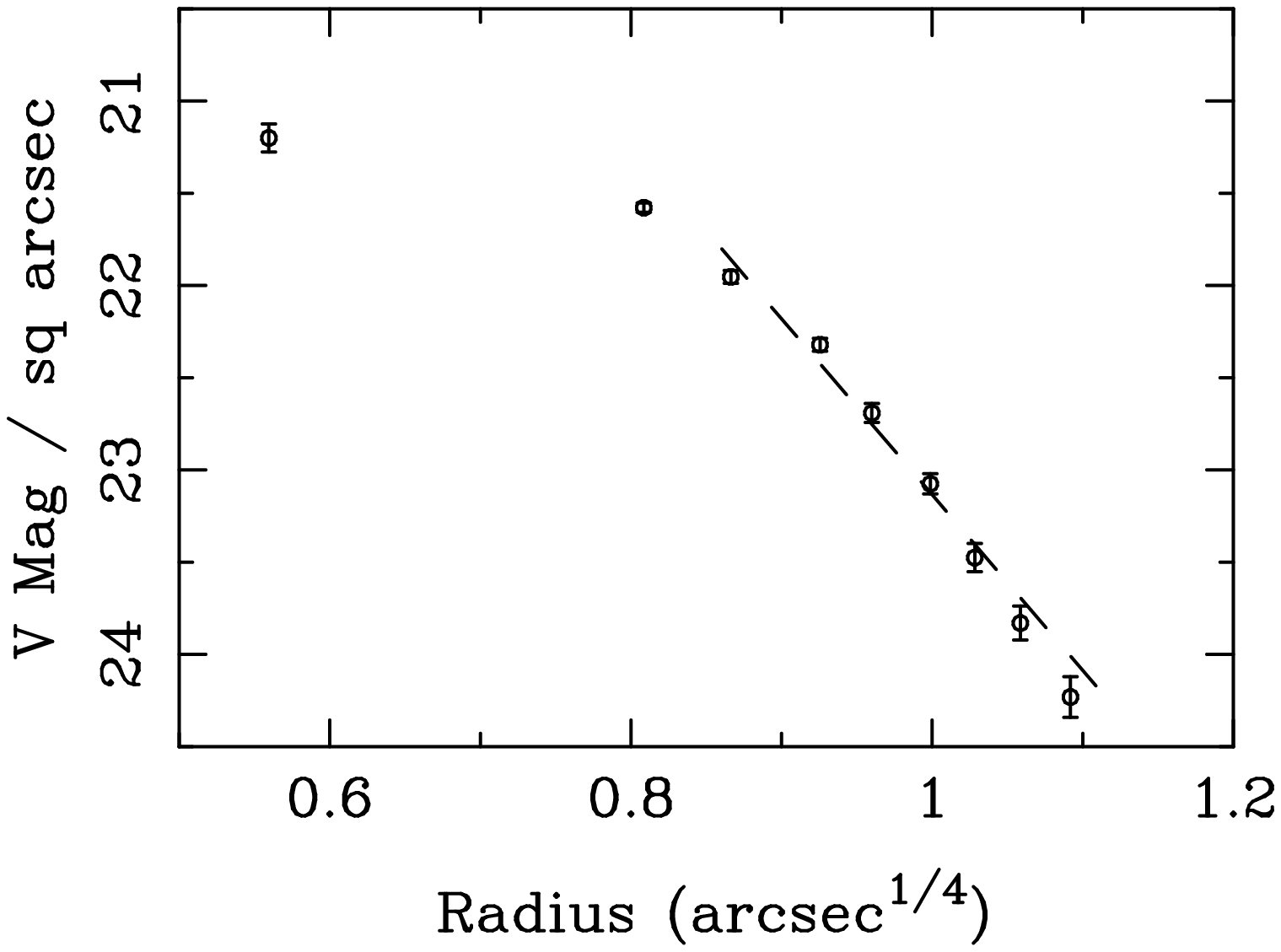} \caption{}\end{figure}
\begin{figure}[h] \figurenum{7}\plotone{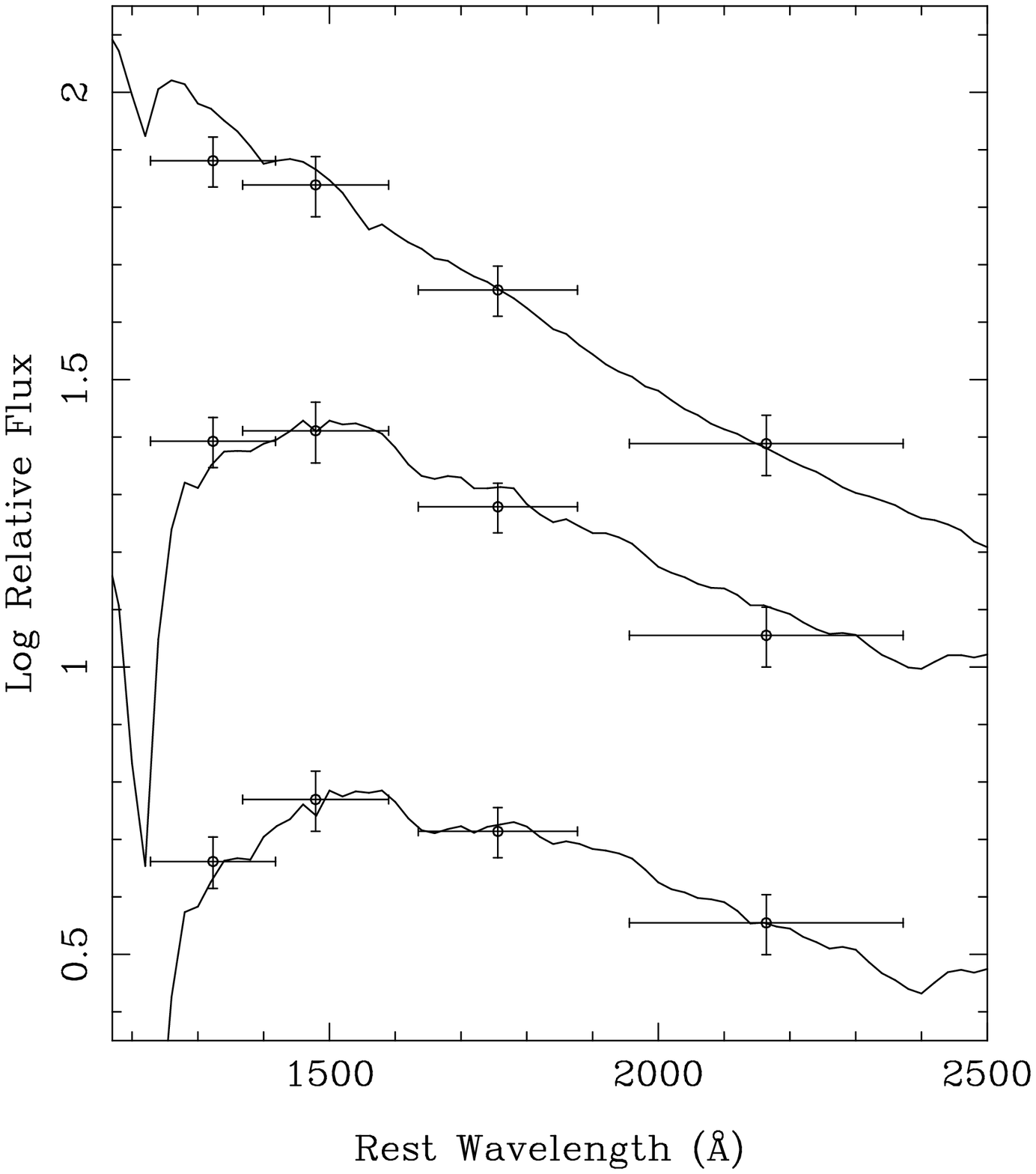} \caption{}\end{figure}
\begin{figure}[h] \figurenum{8}\plotone{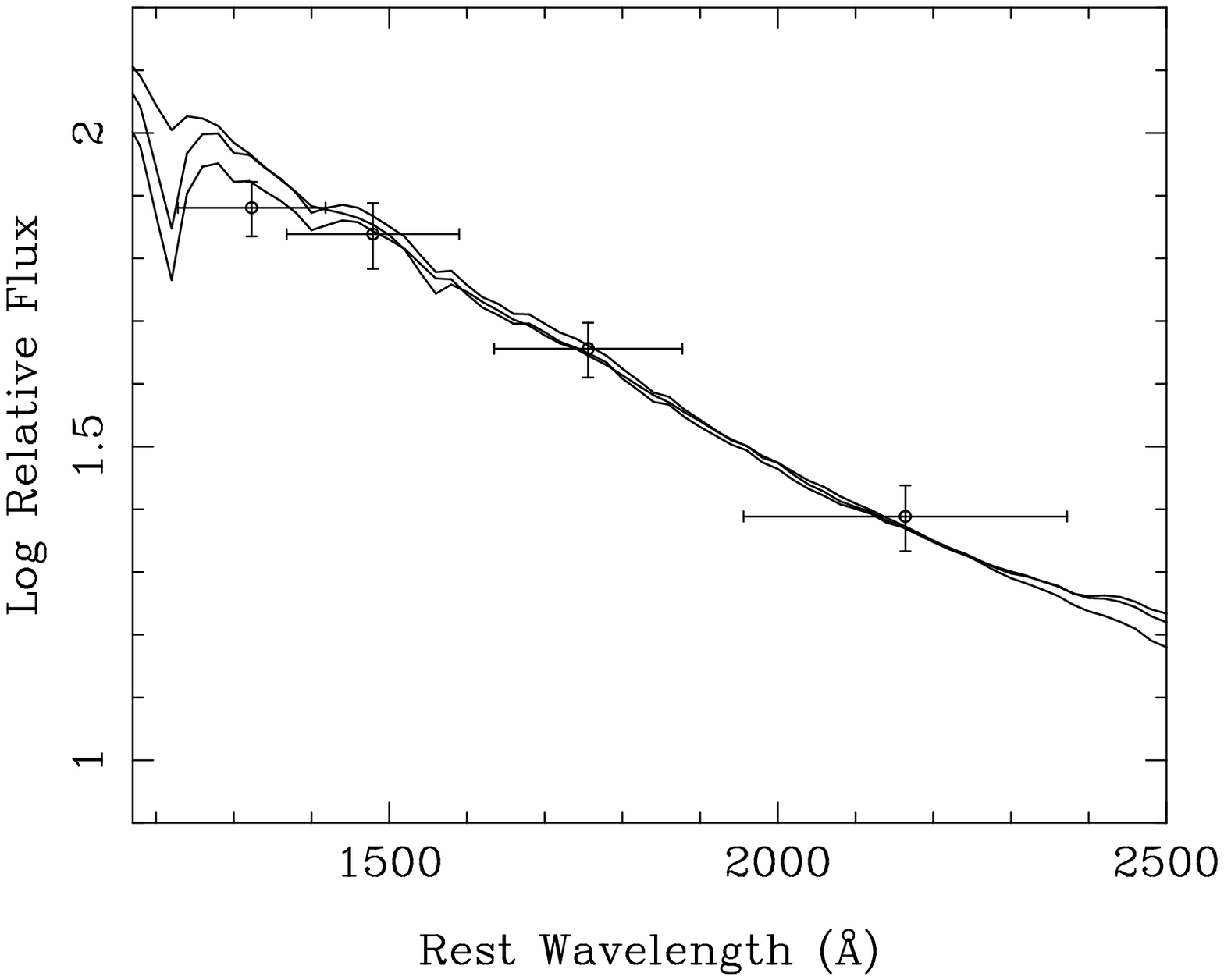} \caption{}\end{figure}
\begin{figure}[h] \figurenum{9}\plotone{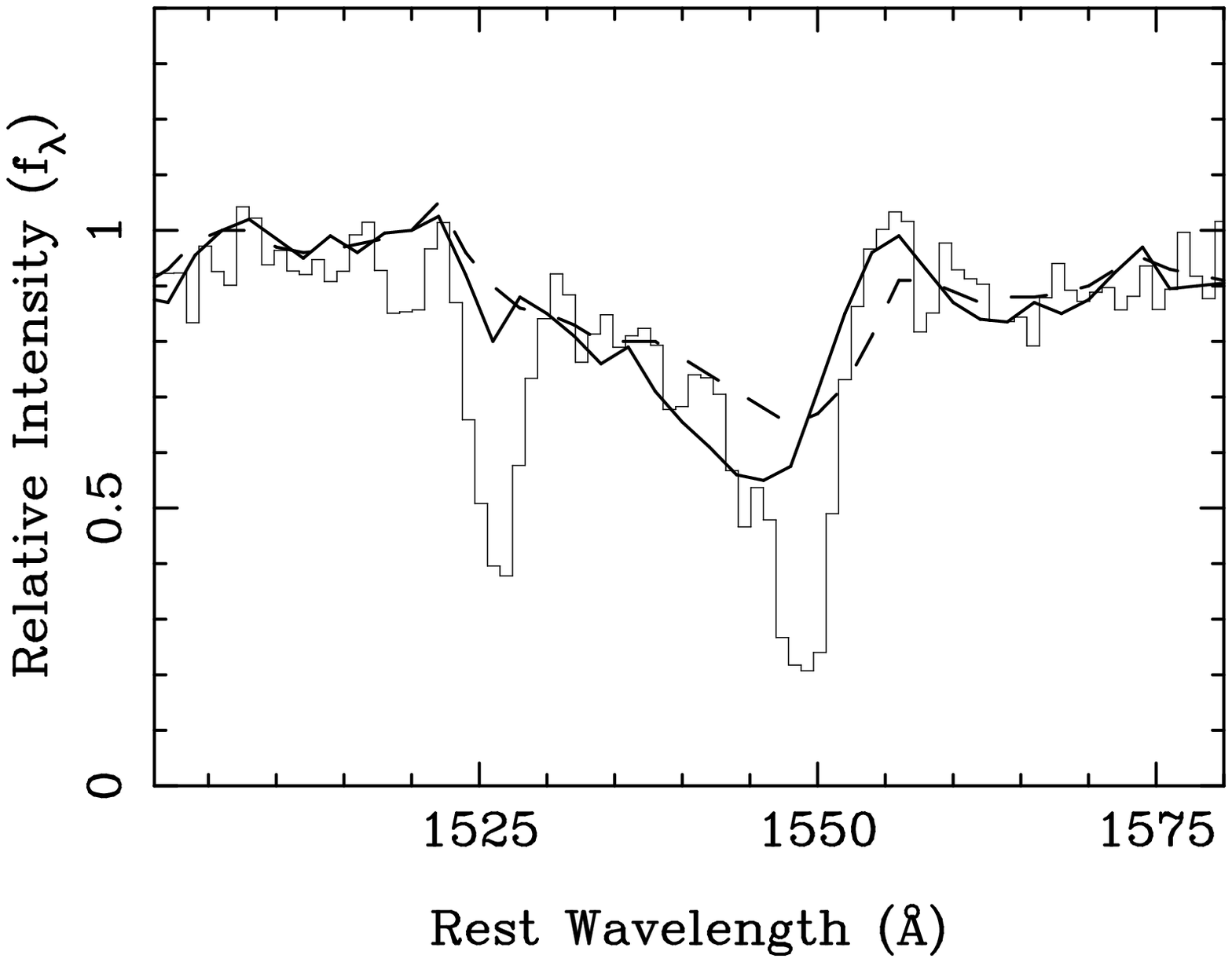} \caption{}\end{figure}
\end{document}